\documentclass{optica-article}

\journal{opticajournal} 

\articletype{Research Article}

\usepackage{lineno}

\begin{document}

\title{Far infrared to terahertz widely tunable narrow linewidth light source via surface-emitting periodically poled thin film lithium niobate waveguides}

\author{Valerie Yoshioka,\authormark{1} Jicheng Jin,\authormark{1} Qiang Lin,\authormark{2} and Bo Zhen \authormark{1,*}}

\address{\authormark{1} Department of Physics and Astronomy, University of Pennsylvania, 209 S 33rd Street, Philadelphia, PA 19104, USA}
\address{\authormark{2} Department of Electrical and Computer Engineering, University of Rochester, 500 Wilson Boulevard, Rochester, NY 14627, USA}

\email{\authormark{*} bozhen@sas.upenn.edu} 

\begin{abstract*} 
Generating widely tunable, continuous wave light at long wavelengths via difference frequency generation (DFG) remains challenging due to high absorption and dispersion. The relatively new platform of thin film lithium niobate enables high-confinement nonlinear waveguides, which could improve efficiency. We simulated DFG in thin film lithium niobate waveguides that are periodically poled for surface emission at 30 THz. Maximum efficiency for a 1 cm device is $9.16 \times 10^{-6}$ W$^{-1}$ assuming $d_{33} = 30$ pm/V. The tuning range within 50$\%$ of efficiency at 30 THz is as wide as 25 THz, from 25 THz to 50 THz.
\end{abstract*}


\section{Introduction}
Long wavelength generation is necessary in many different applications, ranging from bio-sensing \cite{Lucas2013} to spectroscopy \cite{Wulf2010} to material detection \cite{Fan2007}. While many applications can be performed with long wavelength pulses, narrowband continuous wave (CW) light tends to require simpler equipment \cite{Karpowicz2005} and can be more easily used for real-time imaging at room temperature \cite{Zhang2021}. However, efficiently producing widely tunable, narrowband CW light in the far-infrared (FIR) or terahertz (THz) regime remains difficult due to high material absorption and rapidly fluctuating dispersion.

Difference frequency generation (DFG) uses the second-order optical nonlinearity to mix two optical modes that are close in frequency and produce light at the difference frequency, which can be much longer in wavelength. While DFG can produce a wide range of wavelengths, phononic resonances in nonlinear materials introduce high absorption at long wavelengths, often limiting efficient DFG to short propagation lengths. Various methods to mitigate this effect have been proposed and can be separated into two categories: collinear and noncollinear.

In the first category, collinear DFG allows both input optical signals and the DFG signal to travel together along the nonlinear waveguide. To reduce the loss felt by the DFG signal, an adjacent waveguide with lower absorption can be used to guide most of the DFG light \cite{Yang2021}. While the long wavelength DFG signal experiences less loss, it still interacts with the nonlinear material, allowing the process to continue. However, some portion of the DFG light must continue to interact with the optical pump signals in the lossy, nonlinear material for collinear DFG to occur. As such, it experiences exponential loss as it propagates. While this scheme is advantageous for wavelengths with reasonable loss in the nonlinear material, it becomes prohibitively inefficient with higher losses.

In comparison, noncollinear DFG schemes encourage surface emission of long wavelength light to reduce interaction with the lossy material and thus prevent absorption. One of the most common noncollinear techniques is Cherenkov phase matching, which depends on material refractive index. The Cherenkov condition can be summarized by $\phi_C = \arccos\left(\frac{n_{\rm opt}}{n_{\rm FIR}}\right)$, where $\phi_C$ is the emission angle of the DFG light from the sample, $n_{\rm opt}$ is the refractive index at optical wavelengths, and $n_{\rm FIR}$ is the refractive index at the longer DFG wavelength \cite{DeRegis2018-AS}. While this method works effectively in the low THz regime ($<8$ THz) in lithium niobate (LN) due to minimal dispersion, it becomes more difficult in regions where the refractive index is highly dispersive. 

Another method of surface emission interfaces metal antennas with integrated LN waveguides in order to couple out THz light generated from optical rectification of pulsed light \cite{Herter2023}. While antenna design parameters provide significant control over the emitted THz waveform, efficiency is limited since DFG occurs over a short distance, around 200 $\mu$m at most per antenna. As such, implementing a similar scheme with CW DFG would be inefficient and limited by these short scales. 

An alternate approach uses periodic poling to phase match DFG light to radiation modes. Instead of being guided in a waveguide, the DFG light immediately radiates out normal to the waveguide. Analytical derivations of this behavior \cite{Suhara2003}, along with demonstrations of its performance using PPLN \cite{Suizu2006, Hamazaki2022}, have been implemented. However, these approaches were limited by low-confinement, bulk PPLN waveguides \cite{Hamazaki2022, Ding2011}. In recent years, thin film lithium niobate (TFLN) platforms have become commercially available, resulting in more efficient frequency conversion due to higher confinement in integrated waveguides \cite{Chen2022, Boes2023}. 

Here, we focus on the potential for using periodically poled TFLN waveguides as opposed to bulk or diffused waveguides to enhance efficiency and compare performance with more recent developments in long wavelength generation via integrated photonic devices. To our knowledge, no integrated TFLN waveguide has been explored for continuous wave surface emission via periodic poling. In this work, we demonstrate a design for a surface-emitting, periodically poled TFLN device that emits widely tunable, narrowband FIR CW light at room temperature. 

\section{Analytical Model}
In a waveguided DFG process, two overlapping optical modes interact in a nonlinear material, generating a nonlinear polarization that acts as a source for the desired DFG mode. The frequencies of the optical modes, $f_1$ and $f_2$, are chosen such that their difference is the desired output frequency, $f_3 = f_1-f_2$. For x-cut LN waveguides with TE-like optical modes, the electric field can be written as $E_i = A_i e^{i(k_iy-\omega_i t)} \hat{z}$, where $A_i$ is the field amplitude, $k_i = 2 \pi n_{i} / \lambda_i$ is momentum dependent on effective mode index $n_{i}$ at wavelength $\lambda_i$, and $\omega_i = 2\pi f_i$ is angular frequency. The fields travel along the crystal's $\hat{y}$ axis and are directed along its $\hat{z}$ axis. As a result, only the $\chi^{(2)}_{zzz} = 2d_{33}$ term contributes to the nonlinear polarization. Nonlinear polarization can thus be written as: 

\begin{align}
    P_{\rm NL} &= 2 \epsilon_0 \chi^{(2)}_{zzz} E_1 E_2^*\\
    P_{\rm NL} &= 4 \epsilon_0 d_{33} A_1 A_2^* e^{i(k_1-k_2)y-i(\omega_1-\omega_2)t} \hat{z}
\end{align}
A factor of 2 accounts for both permutations of $E_1$ and $E_2$. $P_{\rm NL}$ acts as a source for $E_3$ via the wave equation, resulting in the vector differential equation:
\begin{align}
    (\Delta + k_3^2n_3^2) E_3 &= -k_3^2 \frac{1}{\epsilon_0} P_{\rm NL}
\end{align}
where $\Delta$ is the vector Laplacian and $n_3$ is the refractive index of LN at $\lambda_3$. 

\begin{figure}[ht!]
\centering\includegraphics[width=10cm]{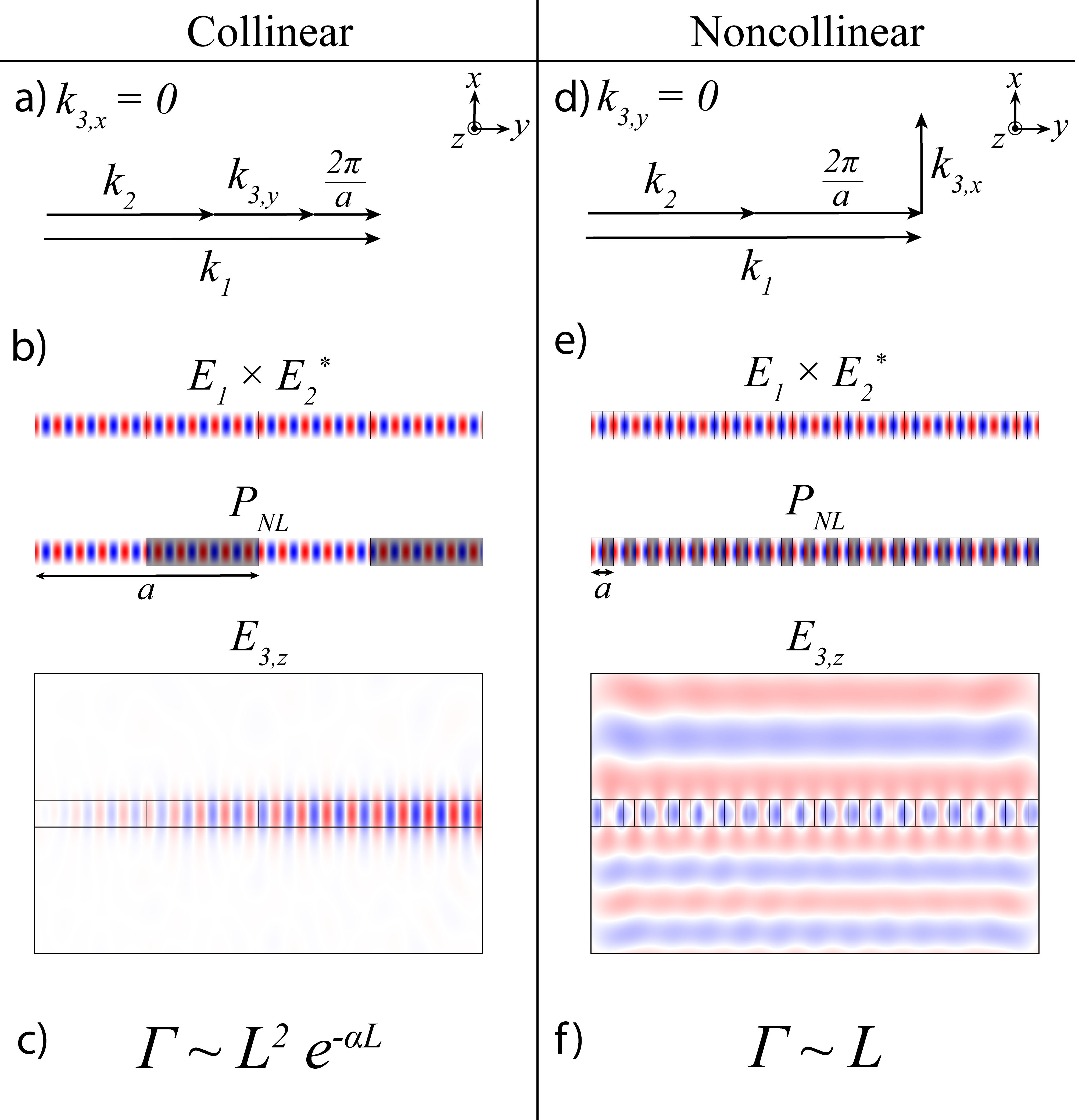}
\caption{DFG with periodic poling for collinear (a-c) and noncollinear (d-f) approaches. a) Phase matching in the collinear scheme; $k_1$ and $k_2$ are the momentum for input optical modes. Poling period $a$ is chosen such that $k_1 - k_2 - k_3 -\frac{2\pi}{a} = 0$. As a result, the $k_{3,x}$ component is set to 0, and $k_3$ is fully directed along the waveguide. b) Nonlinear polarization $E_1 \times E_2^*$ without poling (unshaded) and nonlinear polarization $P_{\rm NL}$ after poling with collinear period $a$ (shaded). The generated output field $E_{3,z}$ is a guided mode. c) Power efficiency $\Gamma$ scales quadratically with length $L$ but decays exponentially with loss $\alpha$. d) Phase matching in the noncollinear scheme. Poling period $a$ is chosen such that $k_1 - k_2 -\frac{2\pi}{a} = 0$, so the $k_{3,y}$ component along the waveguide is set to 0, and $k_{3}$ is fully directed along the $\hat{x}$ direction. e) Nonlinear polarization $E_1 \times E_2^*$ without poling (unshaded) is the same as the collinear case. Below, nonlinear polarization $P_{\rm NL}$ after poling (shaded) is shown with the much shorter noncollinear period $a$ to match the noncollinear condition. $E_{3,z}$ is a radiating mode that emits normally towards both top and bottom of the waveguide. c) Efficiency $\Gamma$ scales linearly with length $L$, but there is no exponential decay with increased length due to loss.}
\end{figure}

For long enough propagation lengths, output power begins to oscillate rather than growing due to phase mismatch, $k_1 - k_2 - k_3 \neq 0$. Quasi-phase matching through ferroelectric periodic poling is often used to counteract this effect. By flipping the material domain and thus the sign of the nonlinear term periodically, phase mismatch can be mitigated and the output power will continue to grow with longer propagation length. This periodic flip can be described by a Fourier series \cite{Boyd}, replacing the constant $d_{33}$ by:
\begin{align}
    d(y) &= d_{33} \sum_m G_m e^{ik_my}\\
    G_m &= \frac{2}{m\pi} \sin\left(\frac{m\pi}{2}\right)\\ k_m &= \frac{2 \pi m}{a}
\end{align}
where $m$ is the index in the Fourier series and $a$ is the poling period. The largest contributing component occurs for $m=1$ and is most frequently used for phase matching. With the $m=1$ poling term, the phase gains an additional term, $\frac{2\pi}{a}$. Using the first and largest poling term, the nonlinear coefficient can be written as $d(y) = \frac{2}{\pi} d_{33} e^{i \frac{2\pi}{a} y}$. Substituting in this nonlinear term adjusts the polarization to become:
\begin{align}
    P_{\rm NL} &= \frac{8}{\pi} \epsilon_0 d_{33} A_1 A_2^* e^{i(k_1-k_2-\frac{2\pi}{a})y-i(\omega_1-\omega_2)t} \hat{z}
\end{align}

Depending on the choice of poling period $a$, the DFG process can be phase matched such that the majority of output DFG light remains in the waveguide as a collinear guided mode or emits via a noncollinear radiated mode. In the collinear case, all $k_n$ are directed along $\hat{y}$, and $a$ is chosen such that $k_1 - k_2 - k_3 - \frac{2\pi}{a} = 0$ (Fig. 1a). The resulting nonlinear polarization gives rise to a guided wave at $\omega_3$ (Fig. 1b). The efficiency scales quadratically with length, but is also exponentially affected by loss (Fig. 1c). 

In comparison, our surface-emitting poling period is chosen to ensure DFG light immediately leaves the waveguide. In this case, $k_{3,y}$ must vanish such that $k_3$ is directed perpendicular to the waveguide. Thus, we choose $a$ such that $k_1 - k_2 - \frac{2\pi}{a} = 0$ exactly (Fig. 1d). In doing so, the phase-matched process only occurs for modes in which $k_{3,y} = 0$. Since $|k_3|$ is defined by the DFG wavelength and material index to be nonzero, and the electric field $E_3$ is polarized in-plane ($\hat{z}$), the DFG light must radiate normal to the surface of the waveguide (Fig. 1e). Efficiency scales linearly with length but does not experience exponential loss with longer propagation lengths since the field quickly radiates away from the lossy material (Fig. 1f). 

Different assumptions are made in solving the wave equation (Eqn. 3) depending on whether the output is a guided mode or radiated light. In the collinear case, guided modes can be treated like plane waves with an additional term that accounts for effective spot area size where all three modes overlap in the nonlinear material \cite{Yang2021}. In the noncollinear case, the output waves radiate in a cylindrical pattern out from the waveguide, so plane wave assumptions break down. The analytical model for noncollinear phase matching in a bulk waveguide has been derived using vector Green functions \cite{Suhara2003}, but our integrated waveguide model includes a few distinctions. 

The first is the inclusion of mode overlap. In a bulk waveguide, the difference between optical mode sizes is negligible. However, modes in integrated thin film LN waveguides are more sensitive to changes in wavelength. Furthermore, while most optical energy is concentrated in the nonlinear waveguide, some portion extends to the cladding and doesn't contribute to DFG; this is especially true for thin, narrow waveguides. When defining overlap for noncollinear DFG, we only account for overlap between the two optical modes in the nonlinear waveguide as the DFG field immediately radiates out of the waveguide. Overlap $\eta$ can thus be written as \cite{Ding2011}:
\begin{align}
    \eta = \frac{\left(\int_{-w/2}^{w/2} dz \int_{-t/2}^{t/2} E_1 E_2 dx\right)^2}{\left(\int_{-\infty}^\infty dz \int_{-\infty}^\infty E_1^2 dx\right)\left(\int_{-\infty}^\infty dz \int_{-\infty}^\infty E_2^2 dx\right)}
\end{align}
where $w$ is the waveguide width and $t$ is the waveguide thickness.  

The second difference in our model is simulated collection efficiency. In comparing our simulation values with the expected efficiency from analytical models, simulated efficiency is affected by the definition of a measurement flux plane (Fig. 2b), through which we measure power outflow. In this way, we include the realistic effects of limited numerical aperture in collection optics, Fresnel coefficients that depend on dispersion, and losses towards the substrate. Note that this factor also changes with waveguide parameters since the energy distribution in the solid angle of the radiated light depends on waveguide parameters \cite{Suhara2003}. 

With these changes, we compare our simulations to the analytical model:
\begin{align}
    \Gamma &= \frac{P_3}{P_1P_2} = ACL\eta\\
    A &= \frac{64 \pi d_{33}^2}{\epsilon_0 c n_1 n_2 \lambda_3^3}
\end{align}
where $\Gamma$ is power efficiency, $P_i$ is the power at frequency $f_i$, $C$ is the collection efficiency factor obtained by comparing simulated output power to our analytical model, $L$ is propagation length, $n_i$ are effective mode indices for respective optical modes, $d_{33}$ is the nonlinear coefficient, and $\eta$ is the mode overlap between the two input optical modes. 

Besides length scaling, noncollinear efficiency differs from collinear efficiency in a few other critical ways. Noncollinear efficiency scales cubically with frequency, as opposed to quadratic scaling for collinear efficiency. As a result, noncollinear efficiency tends to be higher at higher frequencies, and operation bandwidth is asymmetric. Furthermore, in the collinear method, effective spot area is important since higher intensity results in higher conversion efficiency \cite{Yang2021}. In comparison, noncollinear efficiency only depends on normalized optical mode overlap; decreasing mode area does not necessarily improve efficiency. 

Using integrated TFLN waveguides rather than bulk waveguides for noncollinear DFG still has advantages. Tightly confined optical modes in TFLN waveguides ensure most electric field energy is present in the nonlinear material for DFG. Thinner devices also minimize the distance DFG light travels in the nonlinear material before radiating, reducing loss. Aside from potential improvements in efficiency, device size is significantly reduced since TFLN waveguides exhibit better mode confinement. 

\section{Simulation Parameters}
\begin{figure}[ht!]
\centering\includegraphics[width=12cm]{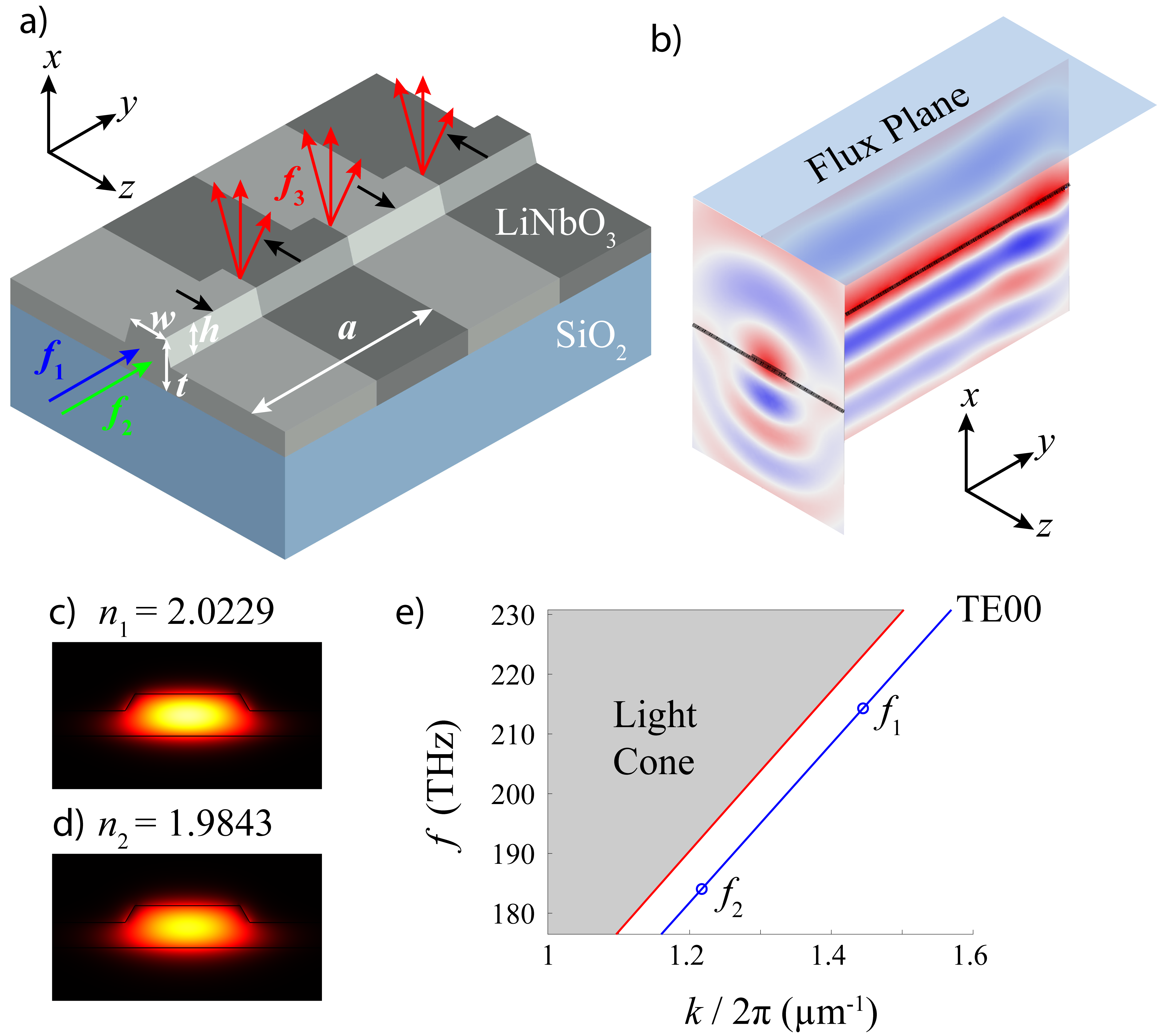}
\caption{a) Diagram of thin film PPLN device. Waveguide parameters include LN thickness $t$, etch depth $h$, and width $w$. Periodic poling with period $a$ is shown with shading; the direction of the $d_{33}$ component is shown with black arrows.  Optical input light signals $f_1$ and $f_2$ are guided into the waveguide. Output light at frequency $f_3$ radiates as a cylindrical wave from the waveguide. b) $E_{3,z}$ radiates from the waveguide as a cylindrical wave; the labeled flux plane is used to define collected output power. c) $|E_1|$ of TE00 mode for $\lambda_1$ = 1.4 $\mu$m with effective index $n_1 = 2.0229$. Waveguide parameters are $t$ = 800 nm, $h$ = 320 nm, and $w$ = 2 $\mu$m; sidewall angle is 60°. d) $|E_2|$ of TE00 mode for $\lambda_2$ = 1.63 $\mu$m with effective index $n_2 = 1.9843$ for the same waveguide. e) Waveguide dispersion for guided fundamental TE00 mode below the light line; $\lambda_1$ = 1.4 $\mu$m ($f_1$ = 214.3 THz) and $\lambda_2$ = 1.63 $\mu$m ($f_2$ = 184 THz) are labeled.}
\end{figure}

Our structure (Fig. 2a) is comprised of a shallow etched TFLN waveguide on top of SiO$_2$ substrate; the top cladding is air. To demonstrate the capability of this technique, we simulated emission at $\lambda_3$ = 10 $\mu$m  ($f_3$ = 30 THz), allowing us to fully capture collection efficiency through 3D simulations with reasonable simulation size and mesh. However, 2D simulations demonstrate that this technique works with even longer wavelengths into the THz regime. For waveguide parameters, we used total LN thickness $t$ = 800 nm, etch depth $h$ = 320 nm, and top width $w$ = 2 $\mu$m with a sidewall angle of 60°. LN waveguides with these dimensions have already been fabricated and poled, proving their experimental feasibility \cite{Wang2023}. Poling period $a$ is set such that the DFG light emits perpendicular to the top surface unless otherwise stated.

In this structure, input wavelengths $\lambda_1$ = 1.4 $\mu$m and $\lambda_2$ = 1.63 $\mu$m have fundamental TE00 modes with effective refractive indices of $n_1 = 2.0229$ and $n_2 = 1.9843$ respectively (Fig. 2c and 2d). In crystal coordinates, TE modes exhibit an electric field pointing in the $\hat{z}$ direction, while the light propagates along the waveguide in the $\hat{y}$ direction. Waveguide dispersion shows that both of these TE00 modes are guided as they lie below the light line (Fig. 2e). 

We used 30 pm/V as the $d_{33}$ value for our DFG simulations. While there is evidence that $d_{33}$ can be much larger around 181-195 pm/V in the THz regime \cite{Boyd1973,Yang2021}, absolute $\chi^{(2)}$ measurements involving frequency mixing between vastly different wavelengths are limited. Furthermore, the $\chi^{(2)}$ values are expected to dramatically fluctuate in the the Reststrahlen regime due to strong dispersion. For LN in the region between 1-100 THz, $\chi^{(2)}_{zzz}$ is expected to vary between values as high as 2000 pm/V to values as low as 2 pm/V \cite{Carnio2017}. However, this dramatic fluctuation has not been rigorously verified experimentally. As such, we set $d_{33}$ to be 30 pm/V as a characteristic value. Note that any changes in this value due to dispersion will cause efficiency to scale proportionally to $d_{33}^2$. If we used $d_{33}$ = 195 pm/V for example, our efficiency would scale up by a factor of 42. In general, additional experimental data on $\chi^{(2)}$ values in the THz regime would be helpful to confirm nonlinear strength at particular wavelengths. 

We used COMSOL to simulate the surface emitting DFG process in periodically poled TFLN waveguides. Input modes were simulated and injected into the waveguide. These optical fields were used to define the nonlinear polarization, which acted as the source for the DFG light at frequency $f_3$. We then simulated the resulting DFG radiation as a cylindrical wave originating from the waveguide (Fig. 2b). To determine expected efficiency, we evaluated the surface integral for power emitted through the flux plane at the top surface. The collection efficiency factor therefore includes losses due to emission towards the substrate and sides, which better reflects the expected experimental efficiency.

\section{Results}
In this scheme, efficiency scales linearly with power (Fig. 3). If the structure is too short, around $L < 10a$, the efficiency is lower than expected due to less uniform power distribution from edge effects. However, as length is increased and edge effects diminish, efficiency trends towards linear increase with propagation length as expected by the analytical model. Reasonable waveguide lengths are much greater than this transition point, so experimental devices will exhibit a linear increase in efficiency with length. For a 1 cm device with 30 THz output, we expect efficiency around $9.16 \times 10^{-6}$ W$^{-1}$. 

\begin{figure}[ht!]
\centering\includegraphics[width=10cm]{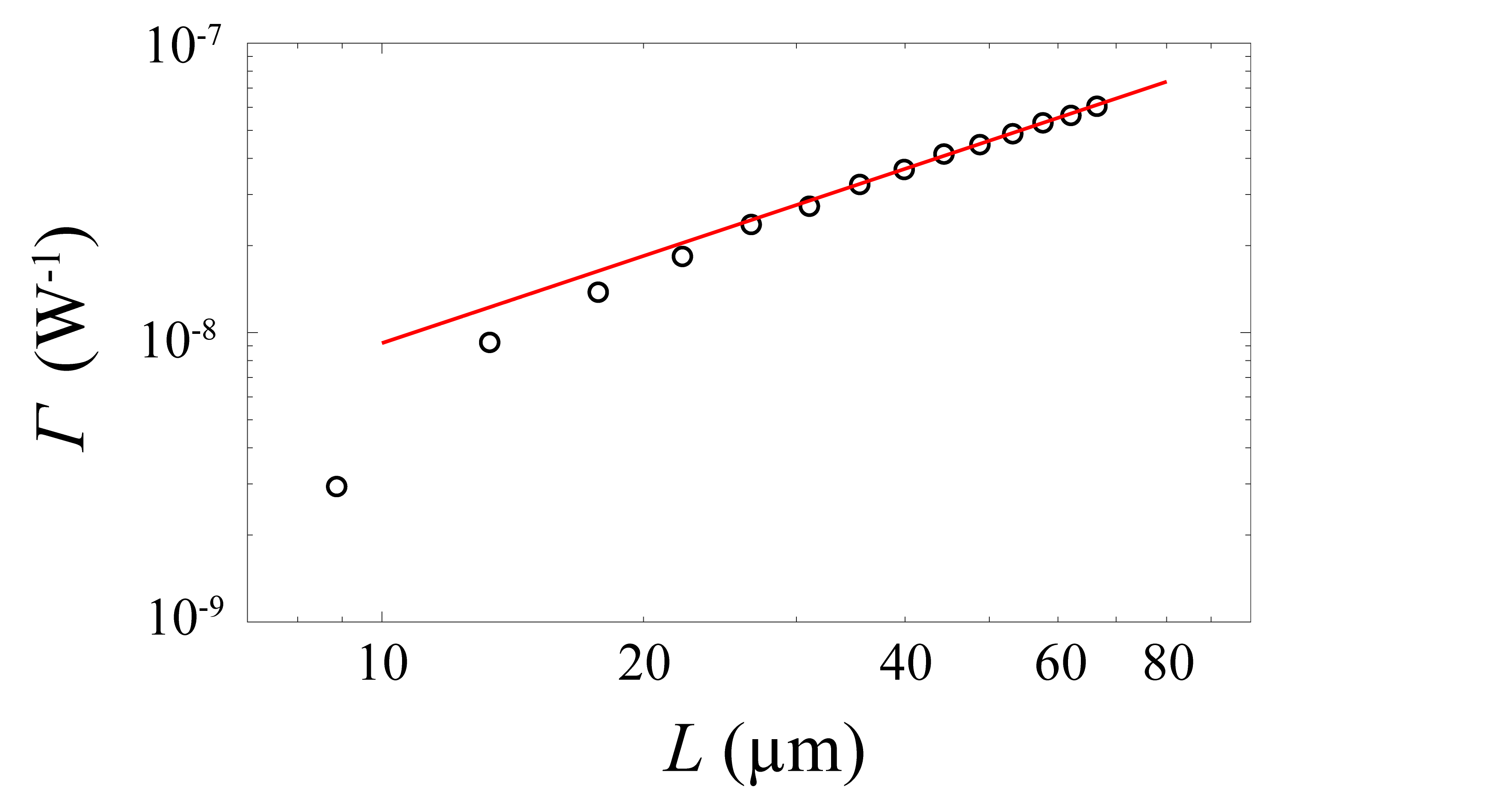}
\caption{Efficiency v. length. In general, efficiency scales linearly with length. The deviation at much smaller lengths is due to edge effects dominating the signal.}
\end{figure}

\begin{figure}[ht!]
\centering\includegraphics[width=13cm]{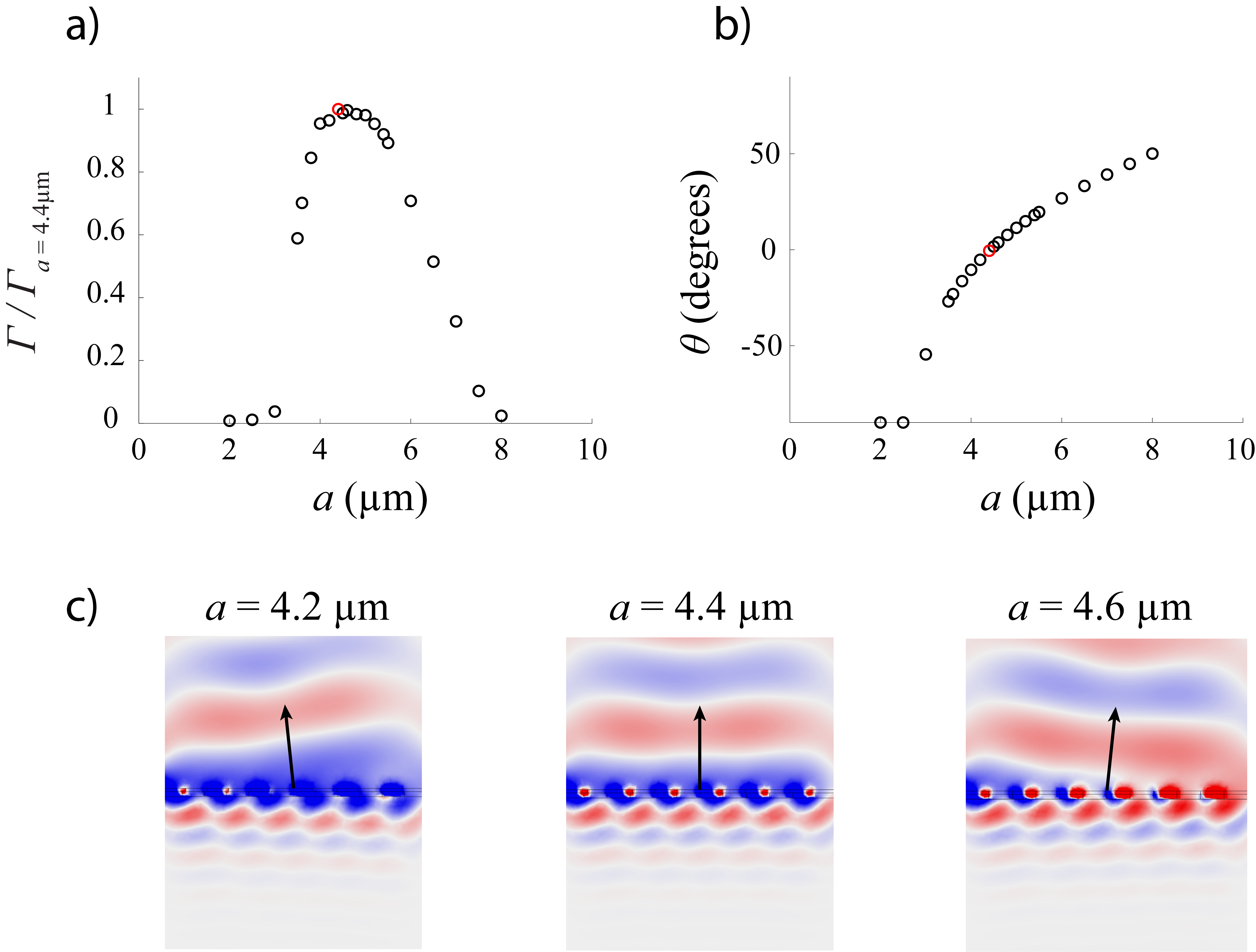}
\caption{DFG emission with varying poling period. a) Normalized efficiency v. poling period; efficiency is normalized by maximum efficiency, which occurs at a poling period of 4.4 $\mu$m (red circle). b) Emission angle v. poling period; the poling period of 4.4 $\mu$m (red circle) has an emission angle around 0°. c) Emission directions for different poling periods $a$ = 4.2 $\mu$m, 4.4 $\mu$m, and 4.6 $\mu$m. Emission angle changes as poling period shifts. Note that emission towards the substrate dies out quickly due to higher loss in oxide at 30 THz.}
\end{figure}

When poling period is adjusted, radiated output light emits at an angle that deviates from fully perpendicular (Fig. 4). This emission angle $\theta$ can be controlled by choosing $a$ such that $k_1 - k_2 - \frac{2\pi}{a} = k_{3,y} = k_3 \sin\theta$. $\theta$ can be positive or negative depending on emission direction but cannot exceed 90° in either direction, limiting the range for poling periods. For our device, poling period must be within the range 2.8 $\mu$m < $a$ < 10.6 $\mu$m to allow emission at 30 THz. 

Efficiency is highest for normal emission ($\theta$ = 0°, $a$ = 4.4 $\mu$m) since our efficiency is defined by evaluating output power through the top surface. Normal emission also reduces loss due to Fresnel reflection coefficients. A steeper angle can result in internal reflection, reducing outgoing light. Fresnel coefficients change according to material dispersion, so critical angles for total internal reflection will depend on the desired central frequency and tuning range.

\begin{figure}[ht!]
\centering\includegraphics[width=10cm]{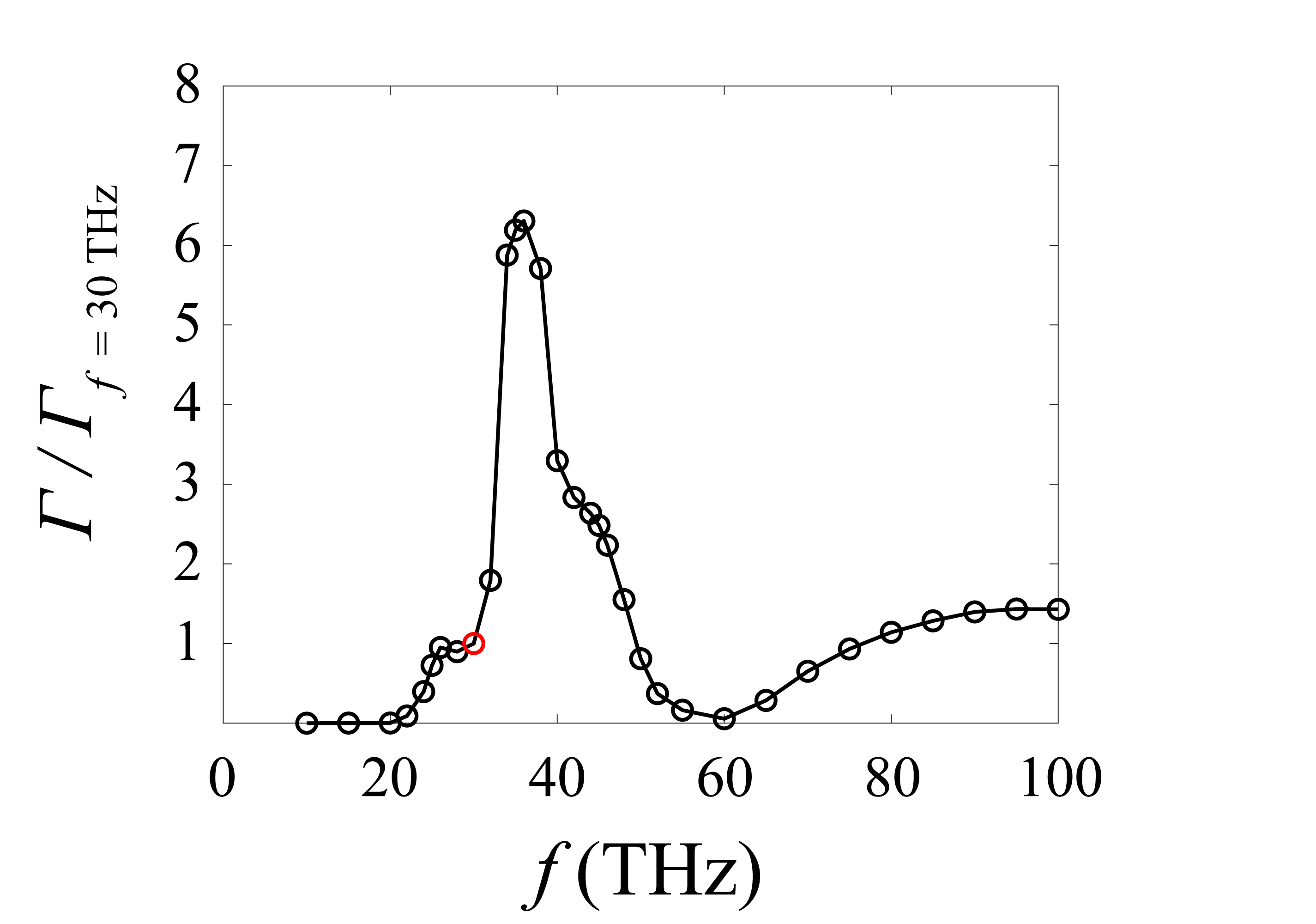}
\caption{Normalized efficiency v. output frequency for a fixed waveguide structure with $a$ = 4.4 $\mu$m. Efficiency is normalized relative to the efficiency at $f_3$ = 30 THz (red circle). Higher frequencies are more efficient due to cubic dependence of efficiency on frequency. The large peak around 35 THz is due to a phonon resonance in the SiO$_2$ substrate, which effectively eliminates substrate loss.}
\end{figure}

To determine bandwidth for a single device, we fixed the poling period to be 4.4 $\mu$m and swept output frequency (Fig. 5). Material dispersion of both LN and the SiO$_2$ substrate was included, resulting in a large peak around 35 THz. At this frequency, phonon resonances in SiO$_2$ affect its refractive index such that imaginary index is larger than real index. The substrate therefore acts like a metal near this frequency, eliminating substrate loss and greatly increasing efficiency. In general, higher frequencies with similar loss tend to be more efficient due to the cubic dependence of efficiency on output frequency $f_3 = c/\lambda_3$. The bandwidth around 30 THz, here defined as the full width at half of the efficiency at 30 THz, ranges from about 25 THz to 50 THz. While asymmetrical, the overall bandwidth spans 25 THz, allowing wide tuning of output wavelengths for different applications.

\section{Comparison to Other Sources}
In collinear DFG, output power scales quadratically with propagation length. However, collinear DFG is exponentially affected by absorption. Even though there are clever approaches to reduce effective loss that the DFG mode experiences, such as guiding it in an alternate material with lower loss \cite{Yang2021}, some portion of the DFG light still needs to travel in the lossy, nonlinear material to benefit from quadratic scaling of collinear scheme, experiencing exponential loss. After a certain length, the exponential absorption is stronger than the quadratic growth with length, and efficiency decreases with longer devices (Fig. 6); this is especially problematic at frequencies above 5 THz where imaginary refractive index $\kappa$ spikes due to phonon resonances in LN. When $\kappa$ is large, the length at which maximum efficiency occurs is too short to efficiently generate long wavelength light. 

\begin{figure}[ht!]
\centering\includegraphics[width=12cm]{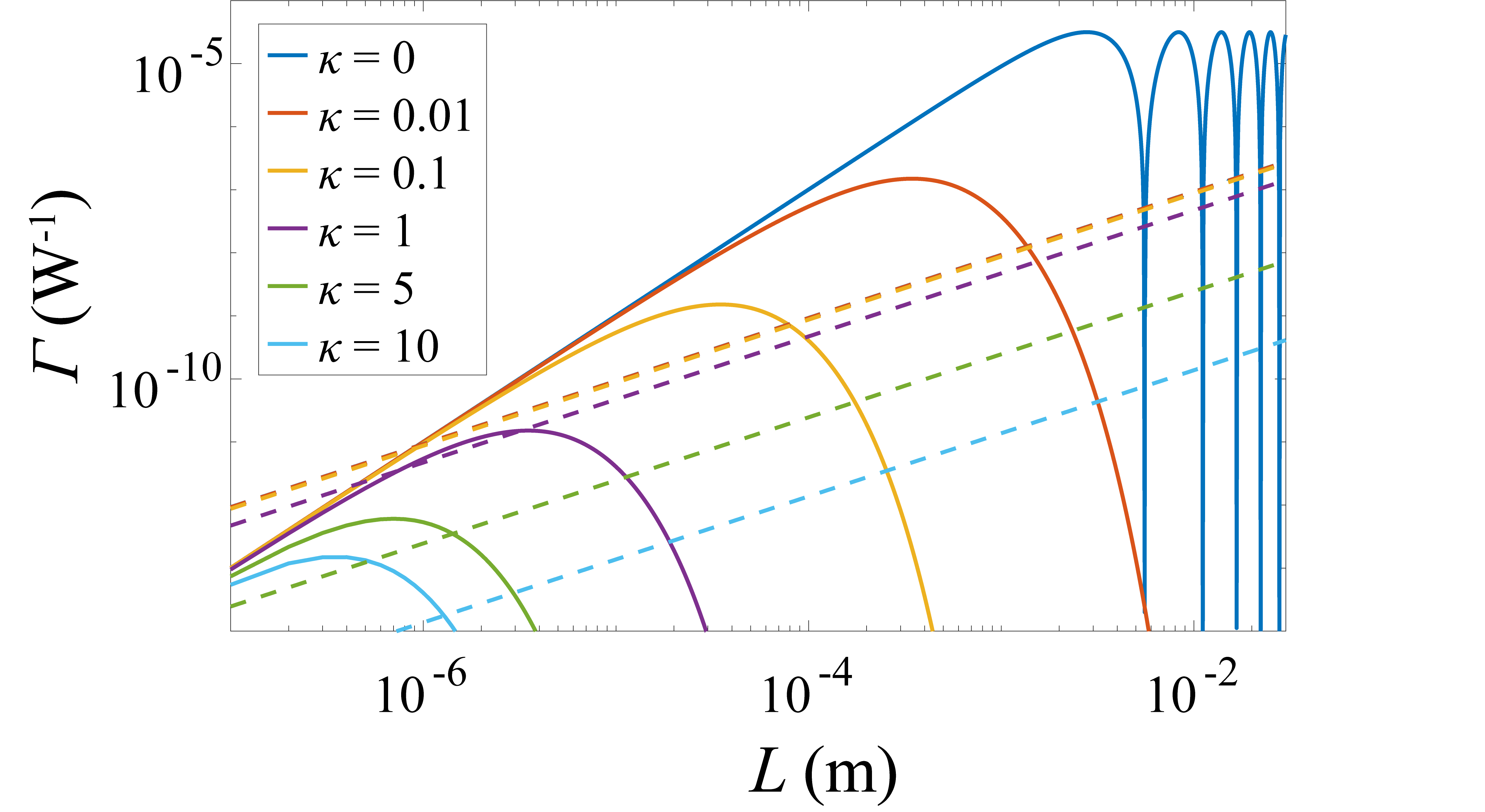}
\caption{Efficiency v. propagation length in both collinear (solid line) and noncollinear (dotted line) schemes for different losses set by imaginary refractive index $\kappa$ (color-coded). Output frequency $f_3$ is 30 THz, and $d_{33}$ is 30 pm/V for both schemes. Differences in noncollinear efficiency at $\kappa = 0, 0.01, 0.1$ are negligible on this scale. For collinear efficiency, phase mismatch affects efficiency through a sinc squared term, resulting in oscillations starting around 5 mm for $\kappa = 0$. For losses of $\kappa = 0.01$ and above, absorption begins to dominate, washing out effects of phase mismatch.}
\end{figure}

In comparison, our noncollinear scheme exhibits efficiency that continues to scale linearly with length regardless of loss since DFG light is immediately emitted and experiences a constant level of absorption. The efficiency per length slightly decreases with higher loss, but propagation length is no longer limited. For long enough propagation lengths, noncollinear devices become more efficient than collinear devices; this crossover length depends on the absorption of the material at the desired wavelength. At 30 THz with $\kappa = 0.0431$ for LN \cite{Carnio2017}, the noncollinear case is more efficient after propagation lengths above 1 mm (Fig. 6), which is reasonable for integrated optics. For $\kappa = 1$, a noncollinear device is more efficient than a collinear one regardless of length. 

Furthermore, collinear schemes are subject to strict phase matching demands, and after a certain length, efficiency is saturated and begins to oscillate. In comparison, noncollinear phase matching requirements are less stringent since a slight deviation in phase matching simply matches the output light to an angled radiation mode, allowing DFG to continue. Efficiency will continue to increase with propagation length, until absorption or depletion of input wavelengths start to become significant. 

In comparison to Cherenkov surface emission schemes, our design is less affected by strong dispersion. In bulk devices, Cherenkov emission is not possible between 10-15 THz or >20 THz since LN refractive index for THz light dips below that for optical light. In waveguide devices, engineering waveguide parameters and using other adjacent materials can tune the effective index for THz light, allowing for Cherenkov emission at otherwise impossible frequencies \cite{Carnio2017}. However, Cherenkov surface emission for continuous wave DFG has only been demonstrated in LN ion-planted slab and channel waveguides with low index contrast \cite{DeRegis2018-AS}, resulting in relatively weak confinement. Furthermore, many devices that rely on Cherenkov emission require silicon prisms on top of the nonlinear material to improve emission towards the top surface since Fresnel coefficients cause internal reflection at the Cherenkov angle \cite{Carnio2017, DeRegis2018-AS}. Since emission angle in our approach relies solely on the choice of poling period, we have a greater degree of freedom in our design to control emission angle in order to maximize efficiency for a central wavelength.

A fair comparison between our simulated efficiency and previous work on surface emission based on periodic poling is difficult since these works used different output wavelengths, nonlinear strength, and underlying assumptions \cite{Suhara2003, Ding2011}. Since many of these parameters are also wavelength dependent, it becomes even more difficult to separate differences in models from how well efficiency is optimized in the waveguide structure. That said, our work uses simulations rather than relying solely on analytical calculations; it thus includes many contributing effects, ranging from material dispersion, mode overlap, limited collection angle, Fresnel coefficients, and any other physical effects that are difficult to express analytically. 

\section{Conclusion}
We have simulated surface emission of 10 $\mu$m CW light using periodically poled TFLN waveguides and demonstrated their potential for efficient generation of narrowband, widely tunable CW light at long wavelengths. While material dispersion still affects device performance at different frequencies, efficiency scales linearly with length and is not exponentially affected by loss. Output power can continually be improved by increasing device length until other effects like pump depletion arise. Furthermore, our device remains reasonably efficient in generating CW light rather than requiring strong pulses, enabling production of narrowband, long wavelength light for applications that require a more precise wavelength or CW light. Frequency tuning also continuously adjusts emission angle, potentially enabling beam steering. In the future, we will consider phase modulating an array of adjacent waveguides to beam steer long wavelength light, allowing even better control over emission angle and tunability.  

\begin{backmatter}

\bmsection{Funding}
This work was partly supported by the U.S. Office of Naval Research (ONR) through grant N00014-21-1-2703, the Army Research Office under award contract W911NF-19-1-0087, and DARPA under agreement HR00112220013.


\end{backmatter}


\bibliography{sample}

\end{document}